# Global Value Chain Linkages and Carbon Emissions embodied in trade, An Evidence from Emerging Economies: Uncovering Connections

[1] Ms. Sakshi Bhayana, [2]Dr.Prof. Biswajit Nag

[1] Indian Institute of Foreign Trade, sakshi_phd2019@iift.edu

Contact No- +917042114900

[2] Indian Institute of Foreign Trade, biswajit@iift.edu

Contact No - +919818200563

**Abstract**

This study explores whether the Global Value Chain(GVC) participation of 16 emerging market economies (EMEs) from 1995 to 2018 in the manufacturing sector leads to a rise in carbon emissions embodied in trade. The study covers the ecological dimension of the Global Value Chain and validates the Pollution Haven Hypothesis in developing nations. To address the problem of cross-sectional dependence, autocorrelation, and heteroscedasticity panels, we estimate the above models using the feasible generalized least squares (FGLS) method. Our findings exhibit a continuous growth in carbon emissions of all the EMEs, there exists a positive association between GVC participation and domestic CO2 emissions embodied in gross exports. Also, EMEs' foreign carbon emissions embodied in gross exports directly correlate with backward GVC Participation, suggesting that the cleaner environment in developed countries comes at the expense of a dirtier environment in developing countries.

Originality: Our research contributes to the sparse theoretical literature on the effect of Value Chain Participation on domestic and foreign emissions embodied in gross exports, especially for EMEs, which is associated with environmental ramifications. This study highlights the connections between decomposed embodied carbon emissions( domestic and foreign components of gross exports)and value chain linkages(importation and exportation channels) specifically for the manufacturing sector of emerging economies. Utilizing emission data from input-output tables enhances the capacity to assess national greenhouse gas emissions and depict the structural links of industrial energy consumption.

Not only this, but we also draw relations between various country-based indicators such as the Environmental stringency index, Manufacturing share, and Economic growth with our environmental-based outcomes and draw policy implications in this context. EMEs may benefit from policies that encourage GVC participation that are environmentally conscious.

**Practical Implications:** Future systems for trading greenhouse gas emissions may take shape, both globally and nationally. It begins by examining the future potential role of emissions trading in combating climate change at internationally. The improvement of climate crisis prevention, with a focus on non-renewable source minimization, carbon footprint reduction and clear indicators of ecologic significance, is an interesting input for manufacturing sector . This paper sheds light on the strategic SDGs for this industry and their relationships by highlighting the interrelationships between the SDGs targeted by the manufacturing sector.

**Research Implications** The research studies the relationship between GVCs and carbon emissions and advocates for the promotion of sustainable development through the reduction of greenhouse gas emissions. By examining the impact of GVC involvement on carbon emissions, policymakers may develop targeted programs to reduce the adverse environmental consequences of global commerce and promote more sustainable manufacturing practices.

I. Introduction

GVC refers to "the devoted useful and operational connections where goods and services are composed, assigned and engrossed on a global scale(Kano, Tsang, & Yeung, 2020). GVCs have also been purported as globally dispersed networks of interconnected, value-added businesses focused on specific products and services, exhibiting households, field, states, and potentially small and medium activity within the international abridgement. are tied together. Participation in the backward GVC demonstrates that a nation's commodities incorporate value added that it has previously purchased from overseas. While forward GVC participation demonstrates that an importer nation does not entirely assimilate an exporting nation's products and instead appear in the exports of the importing nation to third parties. In other words, GVC participation is referred to as "forward" if the supplier is at an initial phase of production and "backward" if the transitional inputs are from the stage of production before that.

According to SDG 13 on Climate Action , it is essential to take immediate action to prevent climate change and its disastrous effects in order to save lives and livelihoods., with growing appreciation and external burden on the environmental blow of GVC activities has prompted GVC partners to check and abode the broader environmental impacts of their product-related activities beyond what they carry out internally. (Gölgeci et al., 2019; Poulsen, Ponte, & listeners, 2016). These developments in GVC's led to the origination of the concept of sustainable environmental upgrading.

Footnotes: Foreign $CO_2$ emissions "$CO_2$ emissions embodied in imported intermediate goods and services that are embodied in a domestic industry's exports"

Domestic $CO_2$ emissions " embodied emissions exports that has been generated anywhere in domestic economy"

Martins, Flávio P., et al.(2023) evaluates the methods in which companies in the cocoa supply chain reveal information related to the Sustainable Development Goals (SDGs) in their reports on sustainability by emphasizing the connections among the SDGs that businesses are aiming to achieve. Improving the prevention of the climate crisis through a focus on carbon footprint reduction, non-renewable resource minimization, and obvious indicators of ecological materiality is an intriguing contribution for businesses.

Ilechukwu, N., & Lahiri, S. (2022)uses annual data from 34 Asian countries for the years 1970–2019 to examine the connection between pollution and international trade. This study estimates a regression model that gives the amounts of trade's impact on pollution, when the underlying model is linear, the net effect is positive (international trade leads to a higher level of emission), but when non-linearities are taken into account, the authors find that the scale and composition effects of pollution are positive, but the technique effect is negative.

Chen, X., & Wang, X. (2017) inspects the viability of item blend as a methodology to convey the low carbon production network under the cap-and-exchange strategy. The investigation results show that the channel structure fundamentally affects both monetary and ecological exhibitions. Profits are higher in an integrated supply chain. Interestingly, a decentralized inventory network has lower fossil fuel byproducts.

The cap-and-exchange strategy has an alternate effect on the monetary and natural exhibitions of the inventory network. Sustainability is an outcome obtained via adjustment of tradeoffs .

For an emerging market economies to participate in global division and cooperation as significant Global Value Chain(GVC), it must drive its focus towards the environmental dimension of sustainability. (Jakob et al 2014 ; S.H. Wang et al 2019). Research has shown that involvement in specialized chains is beneficial for raising indigenous technology and manufacturing levels and is a successful plan for developing nations to develop in the new worldwide pattern (Hou et al 2021). However, it is impossible to deny the fact that trade in goods and services may result in pollution transfer from industrialized to developing nations, increasing the need to reduce carbon emissions in these nations. Environment is deeply embedded in the value chains since the processes of production and distribution require ecological inputs in the form of endowments and energy and in return disturb the environment. The more extended an item's value chains, and the greater its exchange volume, the more noteworthy ecological effects The ecological effects of internationally exchanged products are spread via both forward and backward value chain linkages.

According to Gereffi and Sturgeon (2013), the EMEs have recently grown to be substantial exporters of manufactured goods at all stages of production. EMEs have opened up their economies to foreign commerce dramatically, as seen by their expanding participation in global trade. Second, when the global GDP is calculated using purchasing power parity, the EMEs currently make up strong contribution Third, there have been significant economic reforms in these countries and a shift of resources away from primary sector . Numerous studies have shown that better resource allocation brought about by structural reforms has enhanced their production capacity. Depending on the country, these reallocations have taken different forms (Dabla, Ho, & Kyobe, 2016). Regional differences have influenced growth rates, changes in aggregate productivity, and economic structure in these economies. . On one side EME's are the major players in trading activities contributing towards growth and on the other , their manufacturing hubs are termed as pollution havens

The manufacturing sector, which has seen significant expansion over many years of market-oriented reform, is what is driving emerging enations economic growth. In particular, Kaldor's growth laws emphasizes the superiority of manufacturing over services, with manufacturing serving as the primary driver of expansion due to its greater scale efficiencies, higher income elasticity of demand and potential area for employment. These developing economies are industry hubs for many products like Bangladesh, Vietnam and Indonesia are leaders in textile production, China and India are leading producers of automotive parts , also China is at top position for I-phone assembling .All the tasks associated with manufacturing sector cause pollution , for instance bleaching is associated with textile industry ,batteries, phone casting , fuel systems in automotives etc.

For instance manufacturing policies that has led to transformation include Made in China, Make in India 2015, Activist Industrial Policy in Korea, Export Promotion in Thailand , Poland Innovation ICT driven agenda and many more. However, EME's economic expansion has resulted in high levels of pollution and inefficient use of energy and natural resources throughout production . According to Zhu and Sarkis (2007), a number of institutional actors, including the market, the government, and rival suppliers, had increased ecological pressures on the EME's manufacturing sector. Environmental pollution related issues like disposing off the residual wastes, infiltration of water  is undoubtedly becoming worse in these nations.

A few manufacturing policies that have changed the industry are Made in China, Made in India 2015, Poland's Innovation ICT-driven agenda, Korea's Activist Industrial Policy, Thailand's Export Promotion, and many more. On the other hand, excessive pollution and wasteful use of energy and natural resources during manufacturing are the consequences of EME's economic growth. Zhu and Sarkis (2007) contend, however, that the EME's industrial sector is under increasing ecological pressures from a range of institutional actors, including the market, the government, and rival suppliers. There's no denying that environmental contamination issues, such water intrusion and the removal of remaining garbage, are becoming worse in these nations.

.

## II.    Motivation

The  fragmentation of the production via  GVCs has posed an important challenge to trade-led- economic and green growth hypothesis, however, there are only few studies that have provided some insights on the implications of GVCs on sustainable development goals which covers economic and environmental dimension (Banga, 2014, 2016; Gangnes et al., 2014; Hernández & Pedersen, 2017; Nielsen, 2018).

This study calls out for promotion of sustainable development by lowering greenhouse gas emissions, comprehends the connection between GVCs and carbon emissions. Policymakers may create targeted initiatives to lessen the negative environmental effects of global commerce and encourage more sustainable production methods by analyzing the effect of GVC participation on carbon emissions for manufacturing sector of emerging economies . Policies may be put in place, for instance, to promote the adoption of greener technology and production techniques in GVCs or to provide financial incentives to businesses so they lower their carbon footprint across their supply chains. Policymakers should also endeavor to encourage more accountability and transparency in GVCs, which would help producers, manufacturers, and purchasers better comprehend and manage the environmental impact of trade to attain "low carbon value chain".

In a few studies such as Amiti & Freund (2010), Djankov & Hoekman (2000), Fu & Gong (2011), Gereffi (2009), Jurgens & Krzywdzinski (2009), Koopman, Wang & Wei (2008), Pavlnek, Domanski & Guzik (2009), and Xu & Lu (2009), the role of GVCs and advantages associated with them have been investigated with increased foreign direct inflows, enhanced market share as the main drivers behind considering these EMEs. The EMEs have substantially enhanced their exports of raw materials, intermediate goods, and finished commodities in recent years (Gereffi & Sturgeon, 2013). For instance, China has become increasingly important in the GVCs' import and export activities in recent years. The dynamics of GVCs have changed as a result of the rise of the BRICS nations—Brazil, Russia, India, and China, South Africa—on the international stage(Jangam and Rath B, 2021).In order to validate PHH as it involves the process of " offshoring pollution intensive tasks to emerging market economies

Pan, L., Han, W., Li, Y., & Wu, H. (2022) demonstrate, using a US sample and the legitimacy theory, that stringent environmental regulations constitute an external incentive that puts pressure on businesses to transfer their carbon emissions and establish a substantial positive association between environmental law enforcement and the transfer of carbon emissions using foreign capital and reported emissions reductions as proxies. In addition, when confronted with the power of environmental law, businesses feel pressure to conduct themselves in a legitimate and effective manner. Furthermore, businesses are increasingly inclined to choose emissions transfer over emissions cuts if demand rises in either of these sectors.

### III.     Theoretical Framework

Two contradicting views based on worldwide trade and environmental degradation sway on condition: Some contend that emissions will be undermined as emerging nations will embrace less tough ecological guidelines to increment their worldwide competitiveness, "Race to the bottom" theories, this view is based on Pollution Haven Hypothesis (PHH). For example such hypothesis is tested empirical grounds (Zheng D et al 2017) that for an emerging nation like China laws and regulations inhibit pollution restrict developed nations to transfer their industrial activities to other regions .While other studies contend that exchange globalization could empower more proficient portion of assets, specialized advancement, improved ecological guidelines to meet the prerequisites from the created markets and the selection of corporate sets of accepted rules ("gains from exchange" speculations), this view is based on Porter's Hypothesis.

Lanoie, P., Patry, M., & Lajeunesse, R. (2008) explore the aspects of Porter hypothesis using dynamic dimension as it contends that the hypothesis applies more to industries that produce more pollution intensive and it is more applicable to industries with higher levels of exposure to global competition. For some studies, businesses in more competitive industries are more likely to be motivated to innovate in order to cut costs than businesses in less competitive industries (Reinhardt (2000); Cohen (2003)). This may be especially the case when businesses in developed nations are exposed to external competition as compared to businesses in nations with potentially laxer environmental regulations .There exists a positive and significant correlation between regulation rigor and the likelihood of investments in environmental R & D. Arimur et al.(2007), Popp (2006).

Golgeci et al (2021) emphasizes the significance of emerging market enterprises' strategies, capacities, and cooperative GVC connections to promote the environmental sustainability of GVC and enable the effective adoption of environmental practices in developing markets. According to the World Development Report 2020, one view is that GVCs increase the amount of freight and trash produced compared to conventional trade, which drives firms to pick locations with laxer pollution regulations and other view is based on knowledge flows between firms can enable the spread of more environmentally friendly production techniques throughout a GVC. For instance, green goods such as solar panels, electric cars and wind turbines are produced at lower costs in GVCs and with lower environmental costs of consumption. GVC's are associated with carbon leakage ,environmental impacts are borne through net exports of lower-income economies and imports of higher income economies (Michele D 2020).

Following the most well-known Environmental Kuznet Curve (EKC) hypothesis by Grossman and Krueger 1995, several research have looked at the relationships involving productivity expansion and the ecology. Kozluk & Timiliotis (2016) compared effects of pollutants on gross exports with those on domestic value added embedded in exports using the extended environment policy indicator, which covers BRIICS, as part of a global value chain approach and found no evidence in support of PHH rather discovered that trade specialization is significantly impacted by environmental policies. The beneficial role that environmental laws play in reducing carbon emissions, for BRICS' nations present environmental control measures are effective in meeting their pollution reduction goals. The inverted U-shaped relationship between pollution and income is established in part by environmental legislation. Danish, Ulucak, R., Khan, S. U. D., Baloch, M. A., & Li, N. (2020)

Huang, Y., & Zhang, Y. (2023) discusses how digitalization and GVC's position relate to export-embodied carbon emissions (CEEE) with use of panel data and their findings show that advanced countries are more likely than developing ones to experience a decline in CEEE with digitalization and high-tech industries reach the decline stage of the inverted U-shaped curve before low-tech industries.

D Zhang., H Wang., A Löschel., P Zhou. (2021) proposed a multi-region structural decomposition analysis model that shows GVCs improved in environmental quality and accounted for the entire increase in global emission intensity. Regarding the determinants, the sectoral structure of goods sourced along GVCs and sectoral emission intensity were the main drivers of the global emission intensity improvement.

Fei, R., Pan, A., Wu, X., & Xie, Q. (2020) uses the MRIO model and WIOD data for China, quantifies that the trade of intermediate products was the primary cause of China's export trade's rapidly increasing embodied carbon emissions. Additionally, as participation in the GVC division grew deeper, so did its effect on embodied carbon emissions. China's EC position index increased as GVC participation increased, despite China's role as a "pollution haven" in GVCs. Liu, Cenjie, and Guomei Zhao.(2021), show that involvement in the global value chain significantly reduces the embodied carbon emissions. Wang, X., Sun, X., Oprean-Stan, C., & Chang, T. (2023) evaluates the impact of GVC on carbon emissions embodied in exports by recalculating emissions using multi regional input output table, findings indicate that degree of GVC integration has a non linear connection with pollution.

The majority of studies (Alshehry and Belloumi, 2015, Menyah and ,Pao and Tsai 2011) discovered a correlation between pollution and economic growth. Rajpurohit and Sharma (2021) contend exponential growth and financial development reduce carbon emissions and other study shows that moderate economic growth and financial sector development enhance emissions. (Talah et al., 2021). According to Muhammad, S., & Long, X. (2021), $CO_2$ emissions can be decreased by the interaction between political stability and the rule of law , using data for 65 BRI nations ,foreign direct investment (FDI) has a varying effect on carbon emissions depending on the income level, supporting the "Pollution Haven" and "Pollution Halo" theories. Increased CO2 emissions in lower-middle income countries result from trade openness, which lowers CO2 emissions in high income and low income countries. To reduce CO2 emissions, it is critical to strengthen the enforcement of environmental laws.

Environmental upgrading is envisioned as the method of increasing the ecological consequences of production systems, including production, refining, transit, consumption, and waste disposal or recycling. to Poulsen et. al (2018). Environmental consciousness is needed for designing and running internationally linked supply chain networks and investigates various "thermal" transfer mechanisms, talks about the supply chain's overall carbon footprint, and proposes a preliminary analytical model for calculating carbon emissions via supply chain operations. ( Sundarakani, B., De Souza, R., Goh, M., Wagner, S. M., & Manikandan, S. (2010)).

Footnotes : According to the Porter Hypothesis, stringent environmental laws can promote efficiency and innovations that boost a company's ability to compete internationally.
Environmental Kuznet Curve : There exists am inverted U shaped relation between environmental degradation and economy's growth

## IV. Methodology and Data

To explore the relationship between GVC participation and Carbon emissions , we consider both domestic and foreign emission embodied in gross exports based on Leontief Inverse based on inter country input output table ( source :OECD STATS)`.Our study considers emissions from manufacturing sector of 16 EME's with their yearly data spanning from 1995 to 2018 sourced from OECD i.e. study covers 16 cross-sections and 24 annual time periods(N*T=16*24) as there prevailed paucity of data for manufacturing sector of some developing economies, which comes out to be a challenging aspect of our research. Measures for GVC participation (forward GVC: exportation and backward GVC :importation ) are taken from OECD Tiva database. Both carbon emissions embodied in trade and GVC participation involves the process of creating ICIO tables standardizing the formats and classifications of national Supply and Use tables (SUTs) and Input-Output tables (IOTs). Additionally, bilateral trade statistics in goods and services from international sources are integrated with these tables, and they are balanced within predetermined constraints using official National Accounts by economic activity and National Accounts main aggregates time series (SNA08).(OECD Tiva) Numerous data gaps must be filled using a range of estimation techniques before balancing. For instance emissions are calculated in the following way:-

For C is emissions vector by source industry and country, eB is emissions multiplier matrix and $T$ is a matrix of trade flows with each element being a bilateral trade flow. For example, emissions embodied in exports of product p from country 1 to the rest of the world ,

Where N is the no. of countries and K is the no. of industries

$C_{NK*1} = e\,BT$

The positive sign indicates that GVC participation for EME's is positively related with carbon emissions in manufacturing sector via its importation( metal scraps, plastic, rubber) and exportation activities (related to clothing, bleaching , painting etc.).

Other variables of interest that hold relevance and affect CO2 emissions are discussed below

Trade Openness which reflects the industrial sector's openness to foreign trade was measured using the ratio of total imports and exports to total production value. An increase in the industrial sector's openness in developing economies will lead to higher pollution from carbon emissions if the computed coefficient of trade openness is positive. Conversely, let's say that the predicted coefficient is negative. If so, free trade with other countries will help lower carbon pollution in the industrial sector.

Economic Growth: As per Hamakawa's (2004) findings, policymakers worldwide face a significant challenge in balancing economic growth, energy consumption, and environmental preservation, which forces scientists to focus more on examining intricate connections between the objectives of sustainable development. It has become more crucial than ever to comprehend the connection between environmental degradation and economic expansion (Kasman & Duman, 2015). There has been mixed evidence across countries between income levels and environmental degradation. It is possible that environmental improvement and economic growth will coexist, but doing so will need very careful policymaking and a commitment to producing goods and energy in the most environmentally friendly manner possible.

It is important to consider this variable Value Added by Manufacturing Sector as the manufacturing sector drives economic expansion in developing countries and is responsible for the steady increase in output in these economies. Given that increased production requires increased energy consumption, which raises carbon emissions that harm the environment, this development may not be environmentally friendly, given the aforementioned circumstances. Other one is Stringency index by the OECD criterion, which is ranked from 0 to 6, where 6 is the strictest set of rules. For each underlying policy instrument, stringency is defined as a higher either implicit or explicit price on the applicable ecological harm caused by businesses. The main advantage of the EPS index is that it is an easy-to-use proxy based on actual policy.

To capture the relationship between the growth in GVC Participation and carbon emissions embodied in trade , we estimate the following models for country i, at time t:

$$CO2\ Emissions_{it} = \beta_0 + \beta_1 * GVC_{it} + \beta_2\ Manufacturing\ Value\ Added_{it} + \beta_3\ Trade\ Openness_{it} + \beta_4 * GDP + \beta_5 * Stringency + GVC_{it} * country\ characterstics_{it} + \varepsilon_{it}$$

Dependent Variable is classified in the following ways one is Domestic CO2 emissions embodied in gross exports, by industry i in country / region c to partner country/region p, represents the embodied CO2 emissions in exports that has been generated anywhere in the domestic economy and the other one is Foreign CO2 emissions embodied in gross exports captures the CO2 emissions embodied in imported intermediate goods and services that are embodied in a domestic industry's exports. The emissions can come from any foreign industry upstream in the production chain. Our main explanatory variable Variables of our model is GVC participation that is considered as Forward and Backward :Through access to inputs and information flows, buying intermediate inputs from other nations boosts productivity and output. This indicator is foreign value added embodied in gross exports, also known as Backward GVC Participation. Forward Participation in GVC is the practice of selling domestic intermediates to overseas customers i.e. the domestic value added re exported to third countries .Other independent variables are trade openness (as a percentage of GDP), GDP per Capita , Manufacturing share( as percentage of GDP), Environmental Stringency Index . Also country specific controls which are considered are Population Density , Renewable Energy Consumption and Forest Cover. Variables are described below in Table 1.

**Table I :Variable Description**

| Variable | Representation | Description | Source |
|---|---|---|---|
| Foreign CO2 Emissions embodied in gross Exports(Manufacturing Sector) Tones million | Foreign CO2 | CO2 emissions embodied in imported intermediate goods and services that are embodied in a domestic industry's exports. | OECD Stat |

| Domestic carbon Emissions embodied in gross exports (Manufacturing Sector) Tonnes million | Domestic CO2 | Domestic CO2 By industry i in country c to partner region represents the embodied CO2 emissions in exports that has been generated anywhere in domestic economy | OECD Stat |
|---|---|---|---|
| Economic Growth (current US $) | GDP | GDP per Capita | World Bank Data |
| Forward Global Value Chain Participation ('000 US $) | Forward GVC | Participation in Global Value Chains via domestic value added in partners' exports and final demand (forward linkages) | OECD Tiva |
| Backward Global Value Chain Participation( '000 US $) | Backward GVC | Participation in global value chains (GVCs) via intermediate imports embodied in exports (backward linkages) | OECD Tiva |
| Manufacturing share (% of GDP) | MFG | India's share in Manufacturing as a percentage of GDP | World Bank Data |
| Environmental Stringency Index | ESI | Stringency refers to the degree to which environmental policies put an explicit or implicit price on environmentally harmful behavior.(aggregated using market based policies-taxes, trading schemes and non market based policies- standards and R&D subsidies) | OECD |
| Trade Openness (%of GDP) | TO | Ratio of sum of exports and imports to GDP | World Bank |
| Forest Cover ((% of land area) | FOR_COVER | Percentage of land area under forests | World Bank |
| Renewable Energy Consumption (% of total final energy consumption) | REN_ENERGY_CONS | Percentage of Energy consumption in total final energy consumption | World Bank |
| Population Density | POP_DENSITY | People per sq. km of land area | World Bank |

All the variables are in logarithmic form. To address the problem of cross-sectional dependence, autocorrelation, and heteroscedasticity panels, we estimate the above models using feasible generalized least squares (FGLS) method. FGLS method calculates the covariance matrix and the coefficients of a regression model in the presence of innovations with an unknown covariance matrix. Estimates obtained through FGLS are efficient and consistent . The innovations process is gaussian with mean 0. Innovations are assumed to be heteroskedastic or autocorrelated. Let the regression equation be $y = \beta X_t + \varepsilon_t$.

Here X represents the explanatory variable and y denotes the dependent variable

Let n and p be the sample size and predictors .FGLS estimator is given as follows

Est $\beta = (X^n \varphi^{-1} X)^{-1} (X^n \varphi^{-1} y)$

where estimated φ innovation covariance based on order1 autoregressive model . The estimated coefficient of covariance matrix is denoted as :

$\hat{E}_{fgls} = \sigma^2{}_{fgls} (X^n \varphi^{-1} X)^{-1}$ where

Variance fgls = $y^n [\varphi^{-1} - \varphi^{-1} X (X^n \varphi^{-1} X)^{-1} X^n \varphi^{-1}] \; y/n-p$ .

Estimates obtained through FGLS method are efficient and consistent, if φ is consistent estimator of its population parameter and predictors consists of X are exogenous.

## V. Empirical Results

Summary statistics of variables is given in Appendix . Table 2 gives the evidence of cross-sectional dependence , results show rejection of null hypothesis at 1 % level of significance , hence we can use FGLS Panel Regression method . Results of two models used are stated in Table 5 . The rankings of GVC Participation and carbon emissions embodied in gross exports for our sample of EME's in time periods 1995 and 2018 are given in Table 3 and Table 4.

**Table II :Results from Cross-sectional Dependence**

| Model | Avg Absolute Correlation | Pesaran Statistic |
|---|---|---|
| Dom CO2= f( Forward GVC, TO, MFG, GDP, STR) | 0.490 (0.00) | 13.77 |
| For CO2=f( Backward GVC, TO, MFG, GDP, STR) | 0.45 (0.00) | 12.57 |

Note: We use Pesaran (2004) test for the cross-sectional dependence. H0: cross-sections are not dependent p values are in parentheses.

**Table III: Ranks of Emerging Nations in Forward and Backward GVC participation and Emissions embodied in gross exports in 1995**

| Ranks | Forward Participation | Backward Participation | Foreign Emissions embodied in Gross Exports | Domestic Emissions embodied in Gross Exports |
|---|---|---|---|---|
| 1 | China | South Korea | South Korea | Russia |
| 2 | South Korea | China | China | China |
| 3 | Russia | Thailand | Thailand | South Korea |
| 4 | Indonesia | Czech Republic | Indonesia | India |
| 5 | Poland | Hungary | Russia | South Africa |
| 6 | Czech Republic | Poland | Hungary | Poland |
| 7 | Thailand | Indonesia | South Africa | Indonesia |
| 8 | South Africa | Israel | Czech Republic | Thailand |
| 9 | BRAZIL | Portugal | Poland | Czech Republic |
| 10 | INDIA | Turkey | Israel | Turkey |
| 11 | Turkey | Russia | Portugal | Brazil |
| 12 | Portugal | South Africa | Turkey | Hungary |
| 13 | Israel | Brazil | Brazil | Portugal |
| 14 | Hungary | Slovenia | Slovenia | Israel |
| 15 | Slovenia | India | Vietnam | Vietnam |
| 16 | Vietnam | Vietnam | India | Slovenia |

Note: The ranks reported for GVC Participation are as a proportion of Gros Exports . A low rank implies a country has higher emissions and higher Participation. Forward participation ranks represents exporting nations in a descending order and backward participation ranks represents importing nations in a descending order.

**Table IV: Ranks of Emerging Nations in Forward and Backward GVC participation and Emissions embodied in gross exports in 2018.**

| Ranks from Highest to Lowest in 2018 |
|---|

| Ranks | Forward Participation | Backward Participation | Foreign Emissions embodied in Gross Exports | Domestic Emissions embodied in Gross Exports |
|---|---|---|---|---|
| 1 | China | China | China | China |
| 2 | Russia | South Korea | South Korea | India |
| 3 | South Korea | Thailand | Vietnam | Russia |
| 4 | Indonesia | Hungary | Thailand | South Korea |
| 5 | India | Czech Republic | Poland | South Africa |
| 6 | Czech Republic | Poland | Turkey | Thailand |
| 7 | Brazil | India | Hungary | Vietnam |
| 8 | Poland | Turkey | Indonesia | Indonesia |
| 9 | Thailand | Russia | Czech Republic | Turkey |
| 10 | South Africa | Brazil | Russia | Poland |
| 11 | Turkey | Indonesia | Brazil | Brazil |
| 12 | Hungary | Israel | Portugal | Czech Republic |
| 13 | Israel | Portugal | South Africa | Hungary |
| 14 | Portugal | South Africa | Israel | Portugal |
| 15 | Slovenia | Slovenia | Slovenia | Israel |
| 16 | Vietnam | Vietnam | India | Slovenia |

Note: The ranks reported for GVC Participation are as a proportion of Gross Exports. A low rank implies a country has higher emissions and higher Participation.

**Table V: Regression Results**

| | MODEL 1: Domestic CO2 = f (Forward GVC, (Forward GVC)$^2$, GDP, MFG, STR, TO) | MODEL 2 Foreign CO2 = f (Backward GVC, (Backward GVC)$^2$, GDP, MFG, STR, TO) |
|---|---|---|
| **Explanatory Variables** | **Coefficient** | **Coefficient** |
| Backward GVC | | 0.19** (0.00) |
| (Backward GVC)$^2$ | | 0.24** (0.00) |
| Forward GVC | 0.22** (0.00) | |
| (Forward GVC)$^2$ | 0.04 | |

|  | (0.20) |  |
|---|---|---|
| GDP | -0.01* (0.00) | 0.09** (0.00) |
| MFG | 0.10** (0.00) | 0.27** (0.00) |
| STR | 0.02 (0.24) | 0.01 (0.52) |
| TO | 0.37** (0.00) | 0.59** (0.00) |
| Wald Chi Square | 190 | 197 |
| No. of Cross Sections | 16 | 16 |

Note :** indicates significance at 5 % l.o.s

   *indicates significance at 10 % l.o.s

   Brackets represents lowest level of significance figures (P value)

The positive signs indicate that forward and backward GVC participation are directly related to domestic and foreign CO2 emissions embodied in gross exports respectively for Emerging Economies . From Model 1 ,we infer that one percentage increase in forward GVC Participation causes domestic CO2 emissions to rise by 0.22 percentage . Also for Model 2 , Elasticity of Foreign CO2 emissions with respect to backward GVC Participation is 0.62 units and Squared backward GVC Participation is also significant. There exists a direct relation between Backward GVC Participation and foreign emissions embodied in gross exports, as dependence on imports in Value Chains by emerging economies is driven by various factors like lower degree of innovation technology in EME'S, laxer environmental regulations and excessive energy consumption causes carbon emissions to increase with time. Industries which are engaged in Backward Linkages like construction, transport and Power generation through their intermediate purchases and import penetration invigorate emissions from other blocks. Some core industries like Iron & Steel, Petrochemicals industries which involves combustion processes, although serve as the primary industries, however accounts for the needs of the other downstream sectors. Environment costs are large for EME as production stages with high carbon impacts are pushed into these nations, where environmental stringency is frequently lower for these countries and not significant enough to regulate the emissions.

This direct relationship between backward linkages and foreign emissions embodied in trade emphasis on the fact that emerging countries would hardly benefit from advancements in technology or spillovers. However, manufacturing related activities are pollution-intensive in nature, and have been impacted by the low-end lock-in for a long time due to which GVC processes in advanced countries own core technology and substantial patents on those technologies, and they exploit those technologies to inhibit the development of new knowledge and the improvement of emerging nations' competitiveness. As a result, it has been unable to overcome the technological obstacles to industrial structure upgrades and chain upgrades. Its highly imperative to break this vicious cycle of low end and locked in perspective .

Our results exhibit a positive relationship between Economic Growth and CO2 emissions, mainly due to high "Scale effect" , as increased production and manufacturing activity requires increased inputs and deeper import penetration that involves use of more endowments and natural resources and thereby causing pollution levels to rise (Zhang et al.2012). The pollution is thought to increase to scale with economic growth ,Grossman & Krueger, (1991).On contrary ,due to technique effect with economic growth and rise in the upstream position, the impulse for new technology and innovation causes domestic emissions embodied in gross exports to reduce by less extent .

Quantitative measure of Environment Policy Stringency by OECD is an weighed average of indicators like feed in tariffs for solar and wind , trading schemes , taxes , deposit & refund scheme , standards based on emission limit and R & D subsidies. This index is not a significant contributor towards reduction in CO2 emissions due to many reasons as many advanced countries outsource their pollution intensive task to developing economies like India ,China , Vietnam etc. Also, pollution abatement technique and policy changes involves huge costs in terms of channels of implementation specially for Lower Middle Income or Low Income countries. Calbick and Gunton (2014) demonstrate that environmental management, as measured by the WEF's Global Competitiveness Report Executive Opinion Survey, is negatively correlated with per capita GHG emissions and accounts for about 7% of its cross-sectional variability using cross-sectional data for OECD countries. Trade Openness is also a significant contributor of emissions mainly through importation, as it involves more energy driven processes which is responsible for causing pollution in Emerging Economies .

## VI.    Robustness Checks

To check for the robustness for our results , we have divided our sample into two groups OECD and NON OECD . Table 6 and Table 7 shows the outcomes of emissions through backward and forward participation . Mainly the Results are similar to the previous section , however its noteworthy that elasticity of foreign emissions embodied in gross exports with respect to share of Manufacturing in GDP is of higher for NON OECD group as compared to the OECD group.(Table 7). Also, Stringency index is positively associated with domestic carbon emissions and is not strong enough to abate pollution. We further examine the effect of country based characteristics on carbon emissions embodied in gross exports . The results exhibit that negative impact of forest cover and renewable energy consumption when mediated with Forward and Backward GVC Participation and positive impact with population density when mediated with GVC Participation .

Footnotes : Low end activities : prevents emerging economies from upgrading as they are stuck in lower value added segments of value chain and locked in nations are those which do not have direct access to an ocean.

**Table VI :Results through emissions for Forward GVC Participation for OECD, NON-OECD and all EME's with country characteristics.**

| Dep Variable: Domestic Emissions embodied in Gross Exports | OECD | NON OECD | ALL EME's |
|---|---|---|---|
| Forward GVC | 0.34** (0.05) | 0.30** (0.03) | 0.05** (0.00) |
| GDP | -0.25* (0.05) | -0.20** (0.05) | -0.19** (0.05) |
| MFG | 0.90** (0.05) | 0.10** (0.10) | 0.11 (0.20) |
| STR | 0.05** | 0.01 | 0.02 (0.24) |
| TO | 0.46** (0.00) | 0.50** (0.00) | 0.33** (0.00) |
| Forward GVC*FOR COVER | | | -0.05** (0.00) |
| Forward GVC*REN ENERGY CONS | | | -0.01** (0.00) |
| Forward GVC *POP DENSITY | | | 0.02** (0.04) |
| Wald Chi Square | 103.47 | 189.17 | 390 |
| No. of Observations | 216 | 168 | 384 |
| No. of Cross -Sections | 9 | 7 | 16 |

Note :** indicates significance at 5 % l.o.s

**Table VII: Results through emissions for Backward GVC Participation for OECD, NON-OECD and all EME's with country characteristics**

| Dep Variable: Foreign Emissions embodied in Gross Exports | OECD | NON OECD | ALL EME's |
|---|---|---|---|
| Backward GVC | 0.21** (0.00) | 0.5** (0.00) | 0.37** (0.00) |
| GDP | 0.05** (0.00) | 0.09** (0.00) | 0.13** (0.00) |
| MFG | 0.29* (0.10) | 0.51** (0.00) | 0.54** (0.00) |
| STR | 0.07** (0.02) | -0.03 (0.20) | 0.02 (0.35) |
| TO | 0.15 (0.10) | 0.41** (0.00) | 0.51** (0.00) |
| Backward GVC*FOR COVER | | | -0.02** (0.00) |
| Backward GVC*REN ENERGY CONS | | | 0.03** (0.00) |
| Backward GVC *POP DENSITY | | | 0.04** (0.00) |
| Wald Chi Square | 147.56 | 157 | 496 |
| No. of Observations | 216 | 168 | 384 |
| No. of Cross -Sections | 9 | 7 | 16 |

Note :** indicates significance at 5 % l.o.s

*indicates significance at 10 % l.o.s

Brackets represents lowest level of significance figures (P value)

Another set of robustness check includes the outcomes of emissions through backward and forward participation with fixed time periods so as to justify for the dimensions of our panel data (N * T). Results exhibit direct relation between GVC participation and emissions and other significant regressors are trade openness , country's economic growth and its share in manufacturing sector. (Table 8)

Also, another approach involves dynamic panel models , based on inclusion of lagged dependent variable as one of the regressor and using lagged differenced GVC participation (main explanator variable) as the instrument( Anderson Hsiao , 1981 ) so as to resolve issues like endogeneity and ensures consistent results .This method uses instrument variable and first differenced explanatory variables. Results shows that lagged carbon emissions are significantly contributing towards the current emissions. Other variables which act as major contributors responsible for emissions are forward and backward GVC participation , manufacturing share , trade openness of emerging nations. ( Table 9).

**Table VIII: Panel Data Results with Fixed Time Effects**

| Dep Var: Domestic Emissions | Coefficient | Dep Var: Foreign Emissions | Coefficient |
|---|---|---|---|
| Manufacturing share | 0.0512* (0.06) | Manufacturing share | 0.0345* (0.09) |
| GDP Per Capita | 0.0567** (0.04) | GDP Per Capita | 0.3779** (0.00) |
| Trade openness | 0.0324** (0.02) | Trade openness | 0.2341** (0.04) |
| Forward Participation | 0.2524** (0.00) | Backward Participation | 0.6944** (0.00) |
| Stringency Index | 0.0173 (0.56) | Stringency Index | 0.0082 (0.856) |
| Wald Chi Square | 500.382 | Wald Chi Square | 480.654 |
| Fixed Time Effects | Yes | Fixed Time Effects | Yes |
| No. of Observations | 384 | No. of Observations | 384 |
| No. of Cross sections | 16 | No. of Cross sections | 16 |
| No of time periods | 24 | No of time periods | 24 |

Note :* *indicates significance at 5 % l.o.s

 * indicates significance at 10 % l.o.s

 Brackets represents lowest level of significance figures(P value).

**Table IX: Results through emissions for Backward GVC Participation using Dynamic Panel Model with Instrument Variable**

| Dep Var :Domestic Emissions | Coefficient | Dep Var: Foreign Emissions | Coefficient |
|---|---|---|---|
| Domestic Emissions | | Foreign Emissions | |
| Lagged | 0.4466** (0.00) | Lagged | 0.5773** (0.00) |
| Manufacturing share | | Manufacturing share | |
| First Difference | 0.01438* (0.10) | First Difference | 0.0112* (0.09) |
| GDP Per Capita | | GDP Per Capita | |
| First Difference | 0.0198 (0.15) | First Difference | 0.0231* (0.10) |
| Trade openness | | Trade openness | |
| First Difference | 0.0658** (0.00) | First Difference | 0.7431** (0.00) |
| Forward Participation | | Backward Participation | |
| First Difference | 0.15** (0.00) | First Difference | 019 * (0.06) |
| Stringency Index | | Stringency Index | |
| First Difference | 0.0031 (0.82) | First Difference | 0.0023 (0.72) |
| Constant | 0.0004 (0.00) | Constant | 0.0076 (0.00) |

| Wald Chi Square | 97.12 | | 99.45 |
|---|---|---|---|
| Overall R square | 0.9432 | | 0.9221 |
| No. of Observations | 384 | | 384 |

Note :* *indicates significance at 5 % l.o.s

   * indicates significance at 10 % l.o.s

  Brackets represents lowest level of significance figures(P value).

## VII.    Conclusion

This paper scans through the question whether GVC Participation increases carbon emissions embodied in gross exports ?Through a study of Panel of 16 emerging market economies (EMEs) spanning through time 1995 to 2018. To examine this research question, study incorporates the following steps To examine this research question, study incorporates the following steps

First, the study inspects the patterns of domestic and foreign emissions embodied in gross exports and GVC participation by considering both forward and backward Linkage an Second, we examine whether participation in GVCs through forward and backward linkages causes emissions to rise using panel FGLS technique. Third, several country specific characteristics mediated with GVC Participation are analyzed. Finally, to capture the degree of heterogeneity in terms of extent of emissions and robustness of our results, we categorize our sample of EME'S into two groups :OECD and NON OECD. Our findings provide the following insights, China and South Korea have been the major participants in GVC's Also , China and Russia are the highest contributors towards generating foreign and domestic emissions embodied in trade Second, our empirical findings reveal that both forward and backward participation in GVCs have significantly contributed towards the carbon emissions embodied in trade. Backward GVC linkages exhibit a strong relation with foreign emissions embodied in trade, validating Pollution Haven Hypothesis as advanced economies are outsourcing pollution intensive activities to emerging economies ,which have less stringent environment regulations. Third, we find the control variables such as Population density and renewable energy consumption are mediating the emissions in these countries.

## VIII. Policy Implications

With the proliferation of GVCs has presented emerging nations with possibilities to combine into the worldwide economic system, which has had a major effect on economy and environment. In today's scenario one of the primary obstacle for EME's is despite having to make difficult decisions about the pattern and processes of production fragmentation due to environmental limits, can the manufacturing sector maintain its competitiveness, grow its contribution to the national economy? There are commercially viable examples of the possibilities that low-carbon technology offers, but there are also trade-offs, including perhaps lower profit margins and suppressed demand as the industrial structure changes.(Sajid et al.,2020) How can these nations reconcile the seemingly complex relation between sustainability, GVC participation and growth.

Participation in GVCs also adds to the change in lifestyle. With the right blend of programs, including labor skill development, strict regulatory issues, and regulatory assistance for SMEs, many of these costs can be reduced.(Mitra et al.2020). Direct emission intensity reductions should be achieved through identification of net emitting and important net forward and backward blocks and by devising strategies in the targeted domain. EME'S should take advantage of its enormous spatial potential for industrial access to less expensive alternative energy sources. .EME's must act with prudence and take initiatives which aids pollution reduction like minimizing the amount of paper and plastic in packaging and employing cleaner recycling technology top priorities. The dissemination of more environmentally friendly production methods throughout a GVC can be facilitated via knowledge spillovers between businesses. The environmental costs of consumption are reduced when green products like solar panels, electric vehicles, and wind turbines are produced at GVCs at cheaper costs.(World Development Report 2020).Effective stringency enforcement to regulate the policies that support reduction in emissions is of primary importance so that EMEs may benefit from policies that encourage GVC participation that are environmentally conscious.

The interaction between trade and environment deserves more investigation in the context of a global emergency marked by several crises, including pollution, climate change, biodiversity loss, and socioeconomic inequality. Even if trade and the environment have complicated relationships, it is crucial to pursue policy consistency in order to address this issue without leaving any scope. Existential threats to future generations rise in the absence of clear paths towards decarbonized economies and creative ways to assess economic performance in order to capture these unsustainable trends.

The "beginning of the end" of the fossil fuel era is heralded by the United Nations Climate Change Conference (COP28), which lays the foundation for a quick, fair and equitable transition supported by significant reductions in emissions and increased funding. It acknowledges the scientific evidence that, in order to keep global warming to 1.5°C, greenhouse gas emissions must be reduced by 43% by 2030 when compared to 2019 levels. However, it also includes measures to increase climate finance, strengthen cooperation between governments and important stakeholders, operationalize the loss and damage fund for developing nations and help countries become more resilient to the effects of climate change so that these nations are prevented from the risk of further becoming pollution havens.(Source: UNFCC)

**Limitations of Study**

It is difficult to collect complete data on emissions and manufacturing activities across several nations and at various production stage. With several production phases and a large number of actors participating, GVCs may be dynamic and complicated. It may be challenging to precisely track emissions and to mark the exact sources of pollution as a result of this complexity. While carbon emissions are the main focus of GVC-emission links, there may be important effects from waste creation, air pollution, and water pollution as well. There exists restricted capacity to handle emissions from upstream and downstream activities: GVC-emission links generally concentrate on emissions from production activities, but emissions from downstream (like transportation and disposal) and upstream (like raw material extraction and processing) activities may also be substantial.

**Way Forward**

The study can further be extended by exploring the implications for other sectors which further add value to the manufacturing sector like services in engineering, management, or logistics that save costs and enhance coordination may all help businesses become more productive. Businesses may also set themselves apart from the competition by delivering services in conjunction with the sale of manufactured goods combining services with goods, or adding them to existing products. Services can also assist businesses in removing unofficial obstacles to entering overseas markets and maintaining sales there. Firms with a higher percentage of in-house services produced may have stylized premia in terms of productivity and exports. The export intensity of businesses increases with the percentage of in-house services. However these activities also contribute to pollution in Global Value Chains (GVCs), which extend beyond manufacturing and may include supply chain management, logistics, and transportation. It is highly important to consider value added servicification in manufacturing so as to capture the broader impact of value chains on emissions. Policy ramifications apply to services in pollution-causing value chains can be in the form of regulation and compliance ,green supply chain management ,circular economy and partnerships and collaboration .

## Data Availability Statement

Data is available at an official website of OECD STAT and WDI.

https://stats.oecd.org/

https://databank.worldbank.org/source/world-development-indicators

## Declaration of Conflicting Interests

The authors declared no potential conflicts of interest with respect to research , authorship and/ or publication of the article .

## Funding

This research received no specific grant from any funding agency in the public, commercial, or not-for-profit sectors.

## Acknowledgement

We would like to thank the reviewers for their thoughtful review of the manuscript. Their valuable suggestions were helpful for improving our manuscript.

## Ethical approval

This is the authors' own original work, which has not been previously published elsewhere.

## Consent to participate

Informed consent was obtained from all individual participants included in the study.

## Consent to publish

Additional informed consent was obtained from all individual participants for whom identifying information is included in this article.

## Competing interests

The authors no competing interests.

**Data Availability Statement**

Data is available at an official website of OECD STAT and WDI.

https://stats.oecd.org/

https://databank.worldbank.org/source/world-development-indicators

https://unfccc.int/process-and-meetings/conferences/un-climate-change-conference-united-arab-emirates-nov/dec-2023/about-cop-28

**Appendix**

Descriptive Statistics

| Variable | Obs | Mean | Std. Dev. | Min | Max |
|---|---|---|---|---|---|
| Forward GVC | 384 | 7.394219 | 0.510157 | 6.16 | 8.83 |
| Backward GVC | 384 | 7.316849 | 0.469769 | 5.59 | 8.53 |
| Domestic Emissions embodied in gross exports | 384 | 1.790544 | 0.585038 | 0.591065 | 3.211468 |
| Foreign Emissions embodied in gross exports | 384 | 1.173082 | 0.456977 | 0.278838 | 2.283301 |
| Forest Cover | 384 | 1.502875 | 0.253286 | 0.810871 | 1.83376 |
| Trade Openness | 384 | 1.8758 | 0.235378 | 1.194114 | 2.225937 |
| Population Density | 384 | 2.049499 | 0.435001 | 0.940322 | 2.75 |
| Stringency Index | 384 | 0.049934 | 0.453227 | -1.22185 | 0.69897 |
| GDP Per Capita | 384 | 4.51045 | 1.329636 | 2.57244 | 7.679029 |
| Manufacturing Share in GDP | 384 | 1.28459 | 0.119931 | 1.014399 | 1.511246 |

**Correlation Matrix**

| | Forward Participation | Manufacturing Value Added | GDP Per Capita | Stringency Index | Trade Openness |
|---|---|---|---|---|---|
| Forward Participation | 1 | | | | |
| Manufacturing Value Added | 0.3139 | 1 | | | |
| GDP Per Capita | -0.0329 | 0.1722 | 1 | | |
| Stringency Index | 0.0722 | 0.3067 | 0.3979 | 1 | |
| Trade Openness | 0.2484 | 0.5437 | 0.1447 | 0.638 | 1 |

|  | Backward Participation | Manufacturing Value Added | GDP Per Capita | Stringency Index | Trade Openness |
|---|---|---|---|---|---|
| Backward Participation | 1 |  |  |  |  |
| Manufacturing Value Added | 0.4915 | 1 |  |  |  |
| GDP Per Capita | 0.2336 | 0.1722 | 1 |  |  |
| Stringency Index | 0.2971 | 0.3067 | 0.3979 | 1 |  |
| Trade Openness | 0.3823 | 0.5437 | 0.1447 | 0.638 | 1 |

Scatter Plots of some sample of emerging economies (GVC participation on Y axis and Carbon Emissions on X axis)

Brazil                                                                India

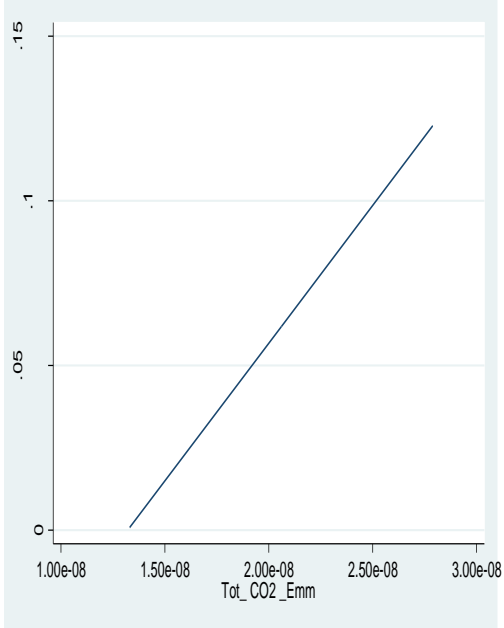
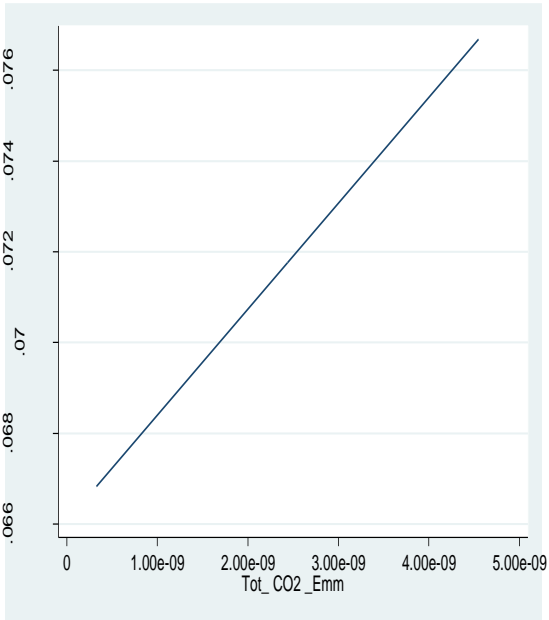

China                                          Russia

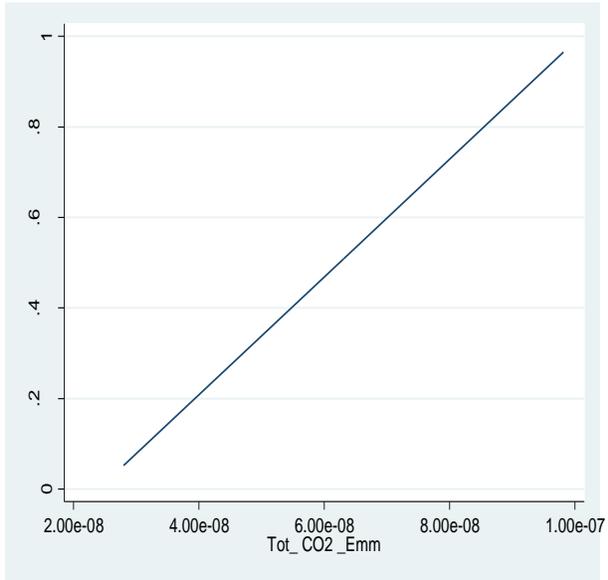
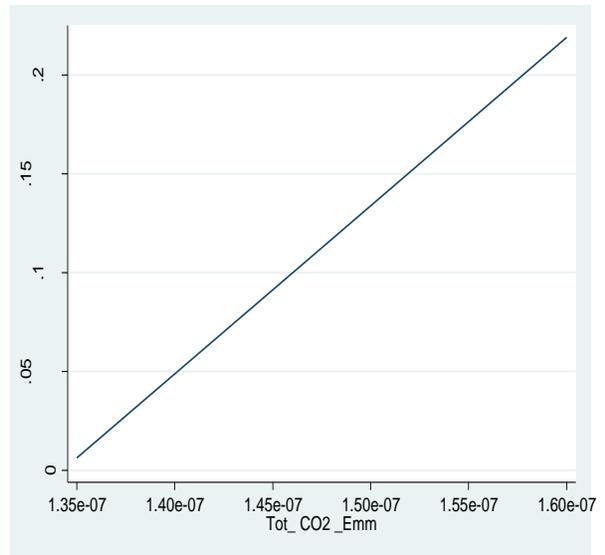

China